\documentclass[12pt]{article}
\usepackage{amsmath,amsfonts,amssymb,amsthm}
\usepackage{mathtext}
\usepackage{epsfig,enumerate}

\newtheorem{stat}{Statement}
\newtheorem*{thm}{Theorem}
\newtheorem{prop}{Propositions}
\newtheorem{lem}{Lemma}
\newtheorem{cor}{Corollary}
\newtheorem{rem}{Remark}

\newcommand{\RR}{\mathbb R}
\newcommand{\ZZ}{\mathbb Z}

\newcommand{\cF}{{\cal F}}
\newcommand{\cD}{{\cal D}}
\newcommand{\cS}{{\cal S}}
\newcommand{\cN}{{\cal N}}
\newcommand{\cI}{{\cal I}}

\renewcommand{\Re}{\mathop{\mathrm{Re}}}
\renewcommand{\Im}{\mathop{\mathrm{Im}}}

\begin{document}

\centerline{\bf P.Grinevich, S.Novikov\footnote {P.G.Grinevich,
 Landau Institute for Theoretical Physics, Moscow, Russia\\
e-mail pgg@landau.ac.ru,\\ S.P.Novikov, University of Maryland
at College Park, USA, and Landau Institute for Theoretical Physics, Moscow,
 Russia\\ e-mail novikov@ipst.umd.edu}$^,$
 \footnote{This work was partially supported by the Russian Foundation
 for Basic Research, grant 11-01-00197-a. The first author was also
 partially supported by  the Russian Federation Government grant
  No~2010-220-01-077, by the program ``Leading scientific schools''
   (grant NSh-4995.2012.1)
and by the program  ``Fundamental problems of nonlinear dynamics'' } }

\vspace{0.5cm}

\centerline{\Large Singular Soliton Operators and Indefinite Metrics}

\vspace{1cm}

{\LARGE Abstract}. {\it We consider  singular real second order
1D Schr\"odinger operators such that all local solutions to the eigenvalue
problems are $x$-meromorphic for all $\lambda$. All algebro-geometrical
potentials (i.e. ``singular finite-gap'' and ``singular solitons'') satisfy to this condition.
A Spectral Theory is constructed for the periodic and rapidly decreasing potentials in the classes
of functions with singularities: The corresponding operators are symmetric
with respect to some
natural indefinite inner product as it was discovered by the present authors.
It has a finite number of negative squares in the both (periodic
and rapidly decreasing) cases.
The time dynamics provided by the KdV hierarchy preserves this number.
The right analog of Fourier Transform on  Riemann Surfaces
with good multiplicative
properties (the R-Fourier Transform) is a partial case of this theory.
The potential has  a pole
in this case  at $x=0$ with asymptotics
 $u\sim g(g+1)/x^2$. Here  $g$ is the
  genus of spectral curve.
}

\pagebreak

{\LARGE {\bf The Main Constructions and Results}}

\vspace{0.1cm}

         {\bf The Fourier Transform and Riemann Surfaces}

\vspace{0.3cm}

Consider real $\infty$-smooth potentials $u(x)$ meromorphic in some small complex
area near the point $x_j\in\RR$ as in
 \cite{GrinNov_UMN_2009,GrinNov_DAN_2011}.
The following  Statement~\ref{stat1} can be easily proved.\footnote {Essentially, this Statement can be 
found in the paper \cite{DuGr}: page 169, Proposition 3.3. The corresponding Corollary for KdV equation 
was proved in this paper for rational potentials only. It is also true for all finite-gap potentials.}
\begin{stat}
\label{stat1}
All solutions to the Sturm-Liouville equations (for all $\lambda$)
$$
L\Psi=-\Psi''+u(x)\Psi=\lambda\Psi
$$
are meromorphic in some small neighbourhoods of the points $x_j$ if and only if
$$
u(x)=n_j(n_j+1)/(x-x_j)^2+\sum\limits_{k=0}^{n_j-1}u_{jk}(x-x_j)^{2k}+ O\bigl((x-x_j)^{2n_j}\bigr)
$$
for some integer  $n_j\in\ZZ$.
There exists a basis of solutions near $x_j$ such that for  $y=x-x_j$ and all $\lambda$
\begin{align}
& \psi_{1j}=1/y^{n_j}+a_1(\lambda)/y^{n_j-2}+ a_2(\lambda)/y^{n_j-4}+\ldots+ a_{n_j}(\lambda)/y^{-n_j}+O(y^{n_j+1}) \\
& \psi_{2j}=y^{n_j+1}+\ldots  \nonumber
\end{align}
\end{stat}
\begin{stat}
All algebrogeometric (AG) potentials satisfy to the conditions of
the Statement~\ref{stat1}.
\end{stat}
By definition for every AG operator $L$ there exists a
linear differential operator $A$ of an odd
order such that $[L,A]=0$. It is well-known that all
eigenfunctions are $x$-meromorphic (see \cite{DMN76}).

We call such potentials $u(x)$ ``singular finite-gap''
if they are periodic in $x$:
$u(x+T)=u(x)$. We call them ``singular solitons'' if $u(x)\rightarrow 0$,
$|x|\rightarrow\infty$.

The simplest examples known in the classical literature are ``the singular solitons'':
$$
u(x)=\frac{n(n+1)}{x^2}, \ \ u(x)=\frac{n(n+1)k^2}{\sinh^2(kx)}
$$
and ``the Lam\'e potentials'' (degenerate and non-degenerate)
$$
u(x)=\frac{n(n+1)k^2}{\sin^2(kx)}, \ \ u(x)=n(n+1)\wp(x)
$$

The ``Dirichlet Problem'' for the real Lam\'e potentials at the
interval $[0,T]$
with real period $T=2\omega$ and imaginary period $T'=2\omega'$
 was studied by Hermit.
No spectral theory for the Lam\'e operators on the whole line $\RR$
 has been studied in the
classical literature.
{\bf We are going to construct a spectral theory for the operator
$L$ in some space of functions on the real line with Indefinite
 Inner Product.}
Let us describe this space.

{\bf Fix a set $X$ of points $x_j\in\RR$, $j=1,\ldots,N$
and numbers $n_j\in\ZZ_+$.} This set should
be finite for the class of rapidly decreasing potentials
$u(x)=O\bigl(1/x^2\bigr)$,
$|x|\rightarrow\infty$. In the periodic case its intersection
with any  period $(x,x+T)$ should be finite\footnote
{There is a very special
interesting case where all $x_j=jT,\ j\in \ZZ$.
In particular, the genus of  spectral curve is exactly
 equal to $n_j=g$ for the famous Lam\'e potentials.}.
For the periodic case we fix also some unitary Bloch multiplier
$\varkappa$,
$|\varkappa|=1$, where
$$
\Psi(x+T)=\varkappa\Psi(x).
$$

We choose class of functions $\cF^{0}_{X}$, $\infty$-smooth outside of
the points $x_j$
(and their periodic shifts), such that near $(x_j)$ we have
\begin{align}
& \Psi(y)+(-1)^{n_j+1}\Psi(-y) = O(y^{n_j+1}), \\
&  y=x-x_j, \ \ j=1,\ldots,N.  \nonumber
\end{align}

The whole space of functions $\cF_{X}\ni\cF^{0}_{X}$ consists
 of functions $\Psi$
with ''principal parts'' $\Phi_j$ at the points $x_j\in X$
\begin{align}
\label{eq3}
& \Phi_j(y)= \sum\limits_{k=0}^{n_j}  a_{jk}/y^{n_j-2k}, \ \ y=x-x_j.\\
& \Psi= \Phi_j +O\bigl(y^{n_j+1}\bigr), \nonumber
\end{align}
so the difference $\Psi-\Phi_j$
satisfies the defining conditions $\cF^{0}_{X}$
locally at $x_j$  for all  $j=1,\ldots,N$. Even more,
this difference has the
order $O(y^{n_j+1})$ at $x_j$.

The standard inner product
\begin{align*}
& <\Psi_1,\Psi_2>=\int\limits_{0}^{T} \Psi_1(x)\overline{\Psi_2}(x) dx, \ \ u(x+T)=u(x), \\
& \mbox{or}\\
& <\Psi_1,\Psi_2>=\int\limits_{-\infty}^{\infty} \Psi_1(x)\overline{\Psi_2}(x) dx, \ \ u(x)\rightarrow 0,
\ \ |x|\rightarrow\infty
\end{align*}
can be extended to the class $\cF_{X}$ by means of the formula
\begin{align}
& <\Psi_1,\Psi_2>=\int\limits_{0}^{T} \Psi_1(x)\overline{\Psi_2}(\bar x) dx, \ \ u(x+T)=u(x), \nonumber \\
& \mbox{or}  \label{eq4} \\
& <\Psi_1,\Psi_2>=\int\limits_{-\infty}^{\infty} \Psi_1(x)\overline{\Psi_2}(\bar x) dx, \ \ u(x)\rightarrow 0,
\ \ |x|\rightarrow\infty \nonumber
\end{align}
(there appears the $\bar x$ instead of $x$ in the second factor
of the integrand compared
with the previous two expressions)
and avoiding  singular points through the complex domain. Our requirements
 imply the following: The
product $ \Psi_1(x)\overline{\Psi_2}(\bar x)$ is $x$-meromorphic. Its
 residues near the
singularities are equal to zero. So our inner product is well-defined
 (but indefinite)\footnote{In our recent papers \cite{GrinNov_UMN_2013}, 
\cite{GrinNov_arxiv_2013} we have shown that this scalar product is 
well-defined on the eigenfunctions of formally symmetric real finite-gap 
operators of arbitrary order and on the eigenfunctions of the 
non-stationary Schr\"odinger operators with one spatial variable.}.

\begin{stat}
\label{stat:st3}
The inner product (\ref{eq4}) is well-defined after this
regularization. It is
indefinite with
exactly  $l_X=\sum\limits_{j=1}^{N} l_{n_j}$ negative squares for each
$\varkappa\in S^1$ where
$$
l_{n_j} = \left[\frac{n_j+1}{2} \right]
$$
\end{stat}

The product  is positive in the subspace  $\cF^{0}_{X}\in\cF_{X}$
 by definition.
Every coefficient $a_{jk}$,
$k=0,\ldots,l_{n_j}-1$ (i.e. corresponding to the negative powers of $y$)
gives exactly one negative square. The positive  powers of $y$ do not destroy
 positivity of inner product.

Detailed proof of the Statement~\ref{stat:st3} is presented
in the Appendix~2.

Let $\Gamma$ be a real hyperelliptic Riemann Surface $w^2=R_{2g+1}(z)$
 of the Bloch-Floquet
 function
$\Psi_{\pm}(x,z)$ in the
``finite-gap'' periodic case. It has exactly two antiholomorphic
involutions $\tau_{\pm}$ where
$z\rightarrow\bar{z}$. We choose $\tau=\tau_+$ such that
the ``infinite cycle''
(i.e. the spectral zone where $z\in\RR$ and $z\rightarrow  +\infty$)
belongs to the  fix point set.
The $z$-poles of $\Psi$ do not depend on $x$.
They form a divisor $\cal D$ consisting of $g$ points. Here $g$ is
the genus of $\Gamma$. By definition the
canonical contour $p_0\in\Gamma$
consist  of all points with
unimodular multipliers $|\varkappa|=1$
(see\cite{GrinNov_UMN_2009}).
In the decomposition theorem below we assume
 that it is nonsingular.
The  $p_0$  is invariant under the action
 of antiinvolution $\tau$. The infinite component of the canonical
 contour contains an infinite point
  $\infty\in \Gamma$.
 The antiinvolution $\tau$ acts trivially on the real part of that
  component.
 By fixing  $\varkappa$
 we get
  a countable set  of
  points $z_q=(\lambda_q(\varkappa),
 \pm)$ in the canonical
  contour. Let us consider the corresponding set of
     functions
 $\Psi_q=\Psi(x,z_q)$. Except of finite  number, all these points
 belong to the infinite component. Our Spectral  Transform
  maps the space of
  $C^{\infty}$-functions on the
  canonical contour (properly decreasing at infinity)  into the
   space of functions $\cF_{X}$ on the real line $R$. It preserves an
  indefinite metric as it
  was proved in
   \cite{GrinNov_UMN_2009}.
   {\bf In the present work we describe the image of this Transform.
    It is the whole space
   $\cF_{X}$.}

     Let us describe first the case of smooth real periodic
    operators:
    The set $X$ is empty.
    All branching points of the Riemann surface are real.

   The divisor $\cal D$ contains
   exactly one simple pole in each finite gap cycle (see \cite{DMN76}.
   The union of all gaps is exactly equal to the fixpoint set of the
   second antiinvolution
   $\tau_-$.

   {\bf The Riemann analog of the Fourier Transform
   (the R-Fourier Transform)
   corresponds to the case of
      real  Riemann surfaces but some branching points may be complex.
       The divisor should be concentrated at
      the infinite point
     ${\cal D}=g \times \infty$.} The Baker--Akhiezer family of functions
     $\psi_x(\gamma)=\Psi(x,\gamma)$ has the best possible
     multiplicative properties in this case: they  are
   similar to the properties of the standard exponential basis
    of the ordinary Fourier
   Transform where the genus zero surface is  $w^2=z$, and the canonical
    contour is an infinite
    cycle over the real positive half-line. For the R-Fourier Transform
     case there exists a singular point $x_j=0$ in the rapidly
      decreasing case,
     and a singular point for every period in the periodic case. It
   is such that $n_j$ is equal to the genus\footnote{In
    the famous cases of the Lam\'e potentials
   there exists only one singular point at the period. We
   investigate in Appendix~3 how many additional singularities
   of the smaller types
   might appear
   in the R-Fourier Transform case.}.
   There exists an operator $R=\partial_x^g+a_1\partial_x^{g-1}+...$
    with coefficients depending
   on $x,y$ only, such that
   $$\Psi(x,P)\Psi(y,P)=R\Psi(x+y,P).$$

We have $a_1=-(\zeta(x+\zeta(y)-\zeta(x+y)))$ for $g=1$ and
 Lam\'e potential (the Hermit case).
It is easy to describe the coefficients for all Riemann surfaces.

   The Riemann analog of Fourier Series with good multiplicative
   properties was developed by
   Krichever and Novikov in the series of works made in the late 1980s.
   They developed the operator
   construction of the bosonic (Polyakov type) string theory for all
   diagrams which are the Riemann
    surfaces of all genera  (see  in the book \cite{String}).
    No analog  of indefinite inner product
   has been  discussed.

\noindent
{\bf Theorem 1.} {\it
Every function $f\in\cF_{X}$ such that $f(x+T)=\varkappa
f(x)$ can be uniquely
presented in the form
$$
f=\sum\limits_q c_q\Psi_q, \ \ \lambda_q=\lambda_q(\varkappa), \ \ c_q=<f,\Psi_q>/< \Psi_q,\Psi_q>.
$$
where $L\Psi_q=\lambda_q\Psi_q$,  $\Psi_q=\Psi(x,z_q)$,
 $z_q=(\lambda_q(\varkappa),+)$ or
$z_q=(\lambda_q(\varkappa),-)$,
and $u(x)$ is a real periodic singular finite-gap potential.
This series converges in the sense that the coefficients
 of the singular parts do
converge (more rapidly than any power), and in some neighborhood
of the points $x_j$ the series $\sum\limits_q(\Psi_q-\Phi_{qj} )c_q$
converges to the corresponding differences $(\Psi-\Phi_{j})$ with all
 derivatives, near every point
$x_j$. Here $\Phi_{qj}=\sum\limits_{q=0}^{n_j-1}a_{(q)jk}/y^{n_j-2k}$,
 $y=x-x_j$, which are the singular
parts of the eigenfunctions $\Psi_q$, at the points $x_j\in X$,
 and  $\Phi_{j}$
is the singular part of $f\in\cF_{X}$ as it was defined above in the
Formula~\ref{eq3}.
}

\noindent
{\bf Theorem 1'.} {\it
Consider a rapidly decreasing potential $u(x)$.
 For every function $f\in\cF_{X}$
decreasing rapidly enough at $|x|\rightarrow\infty$, we have the following
representation
\begin{align*}
& f=\int\limits_{k\in\RR} c_k\Psi_k(x) dk +\sum\limits_{m}d_m\Psi_m, \\
& L\Psi_k=k^2\Psi_k, \ \ L\Psi_m=\lambda_m\Psi_m,
\end{align*}
Here $u(x)=O\bigl(1/x^2\bigr)$ at $|x|\rightarrow\infty$, and
we assume that $u(x)$ is a singular
multisoliton potential.
}

{\bf Remark.} Recently the authors have proved, that decomposition formulas 
from Theorem~1 and  Theorem~1' are valid for all periodic finite-gap 
{\bf real and complex} potentials (they may be regular or singular). 
The proof is based on the reduction to the 
regular complex case and follows the same scheme as used in Theorems 1 and 1'.

Such results for the standard positive Hilbert spaces and regular
 self-adjoint 1D stationary
Schr\"odinger operators where known many years at the folklore level
 (see the formulas and quotations in the
article \cite{DMN76}). For more complicated situation of non-stationary
 1D Schr\"odinger operators and  stationary 2D Schr\"odinger operators
the specific finite-gap formulas and
decomposition theorems on
Riemann surfaces were obtained in the original works
\cite{Krich_FA_1986},
 \cite{Krich_UMN_1989}. In our indefinite case, we use essentially
 the same technique.

{\bf Our program is to extend these results to the whole class
of periodic
and rapidly decreasing infinite-gap real periodic potentials with
 singularities of the type described above.}

By the way, in the work \cite{ArkPorgPol} the ``scattering data''
were constructed for the case
$u(x)=O\bigl(1/x^2\bigr)$ at $|x|\rightarrow\infty$, all $n_j=1$.
Indefinite metric,
 spectral theory
and decomposition of functions were not discussed in this work.

The theory of our functional spaces is based on the solution of
the following
problem:
consider the KdV solutions $u_t=6uu_x-u_{xxx}$ such that $u(x,0)=n(n+1)/x^2$.
 It is well-known, that we can write these solutions in the form
$$
u(x,t)=2\sum\limits_{q=1}^{n(n+1)/2}\frac{1}{(x-x_q(t))^2},
$$
It is easy to see, that  $x_q=a_qt^{1/3}$. How many  of $x_j$-s are real?
Though there is a huge amount of
literature dedicated to the rational, trigonometric
and elliptic solutions to the KdV hierarchy, we could not find the lemma
 below
anywhere. So we
 proved it ourselves.
\begin{stat}
\label{stat4} Exactly $l_n=\left[\frac{n+1}{2}\right]$
poles remain real. This number  is exactly equal to the number of
coefficients $a_{kj}$ at every singular point $x_j$  with $n_j=n$.
\end{stat}

{\bf Remark} The following transformations preserve the
set $\{a_q\}$: $a_q\rightarrow\bar a_q$, $a_q\rightarrow\xi a_q$,
 $\xi^3=1$.

{\bf Proof of Statement~\ref{stat4}.} It is clear, that this problem
is equivalent to the following one:
consider the KdV solutions $u_t=6uu_x-u_{xxx}$ such, that
 $u(x,0)=n(n+1)\wp(x)$, where
$\wp$ is the Weierstrass function, associated with a real
rectangular lattice.
How many  poles remain at the real period  for $t>0$, $t\ll1$?

Let us assume that some generic unitary Bloch multiplier $\varkappa_0$
is fixed.
It follows from the Appendix~2 that the space $\cF_{X}$ has exactly
$l_n=\left[\frac{n+1}{2}\right]$ negative squares for $t=0$.
We use now the  Theorem~1 proved
in the Appendix~1: Any collection of singularities can be approximated
by the image of the Fourier map. Therefore the number of negative squares
is equal to
the number of points $\gamma$ at the canonical contour such that
$\exp(ip(\gamma) T)=\varkappa_0$ and $d\mu(\gamma)/dp(\gamma)<0$.
 Here $d\mu$ is the spectral measure
in the decomposition formula
$$
d\mu=
\frac{(\lambda(\gamma)-\lambda_1)\ldots(\lambda(\gamma)-\lambda_g)}
{2\sqrt{(\lambda(\gamma)-E_0)\ldots(\lambda(\gamma)-E_{2g}))}}
 d\lambda(\gamma).
$$
This number does not depend on $t$. Therefore the number
of negative squares in the metrics in $\cF_{X}$ also does not
depend on $t$.
 For small $t>0$ all
singularities are simple. Therefore the number of negative squares
coincides with
the number of real singular points on the period.
This completes the proof.

Let us point out, that we already proved in the work
\cite{GrinNov_UMN_2009}
that $l'_n\ge l_n$. Here $l'_n$ is the
number of real $a_q$. The $l_n$ is equal to the number
of negative squares
 in the
inner product above for this specific case.
This quantity is time-invariant.
A naive understanding of the opposite inequality $l'_n\le l_n$ is
the following: as
 numerical calculations show,  the points $x_q(t)$ for small
 $t>0$, $t\in\RR$, are
localized approximately in the points of  equilateral triangle
(see Fig 1).

\begin{center}
\mbox{\epsfxsize=12cm \epsffile{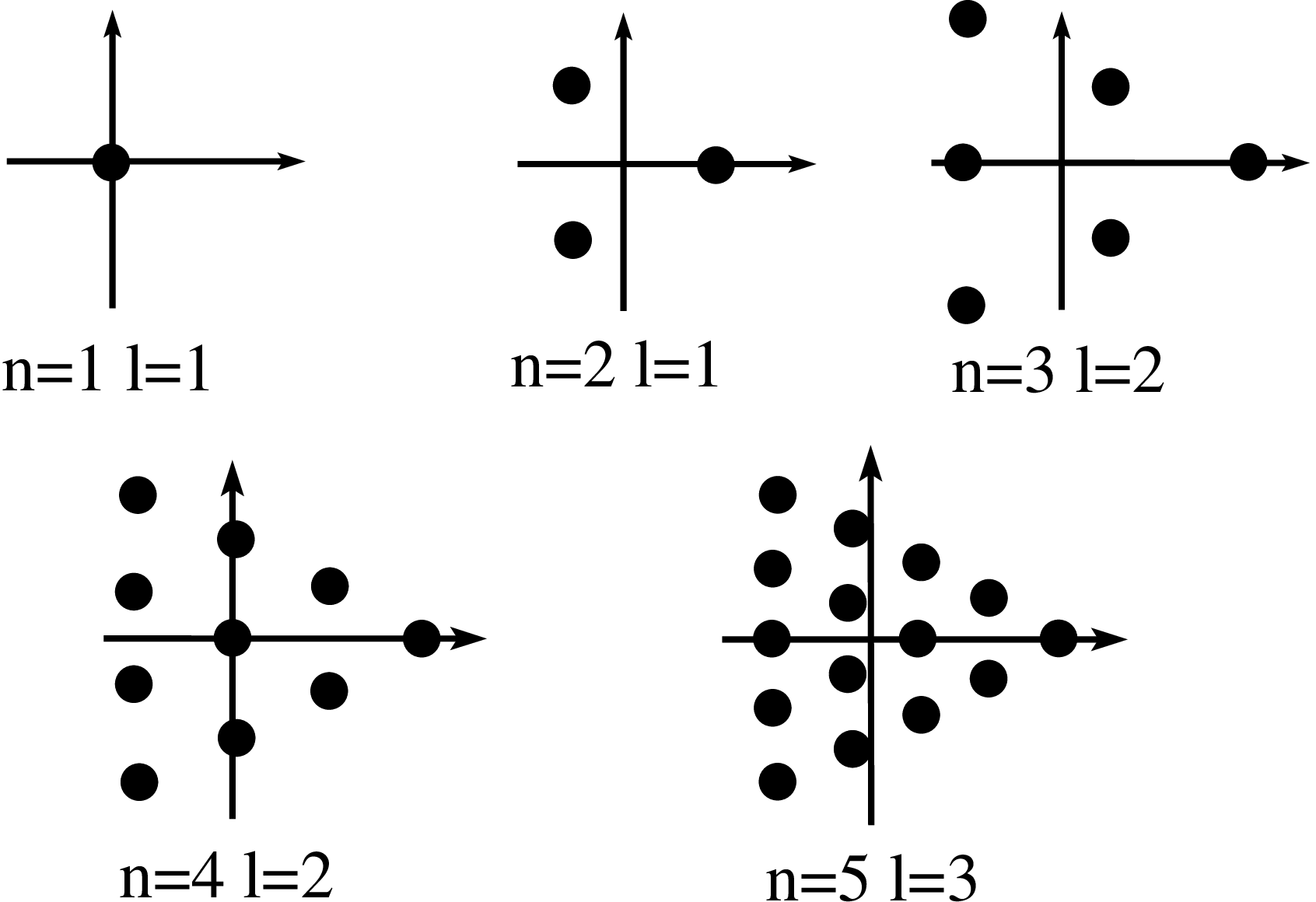}}

Fig 1.

The poles $a_q$ for different values of $n$.
\end{center}

For the case of the ideal equilateral triangle we obviously have $l'_n=l_n=
\left[\frac{n+1}{2}\right]$.

However, in fact, it is slightly perturbed. So we should
have $l'_n\le l_n$ if
perturbation is really small. But the symmetry $a_q\rightarrow\bar a_q$
 keeps all real
points on the real line. So we are done with the really small
 perturbations of the equilateral
triangle. But our perturbation is only numerically small,
 not theoretically. So this
argument is non-rigorous.

{\bf Remark.} The positions of these zeroes were studied numerically
 and analytically in
\cite{ClMan}. However, the problem of calculation of the number
 of real zeroes
 was not discussed in
\cite{ClMan}. It is not clear whether it is possible to obtain
 rigorous proof
of our result based on the estimates from this paper.
The first rigorous proof of the
inequality  $l'_n\le l_n$  was completed with the help of our student
 A.Fetisov. It is different from
the proof presented above.

A non-standard example we obtain for the case of elliptic
function $u(x)=2\wp(x)$
corresponding to the rhombic lattice (see Fig. 2a. Fig. 2b).
 The canonical contour
is connected in this case. It has two singular points.
  The antiinvolution is not
equal to identity at the contour in this case,
so there are no self-adjoint real problems on the real line
for such Riemann surface.
The inner product is always indefinite.  The projection of
the contour to the plane of the spectral parameter contains
a complex part,
so the spectrum of the operator is complex for such real
singular finite-gap potential.

\vspace{1cm}

\begin{center}
\mbox{\epsfxsize=6cm \epsffile{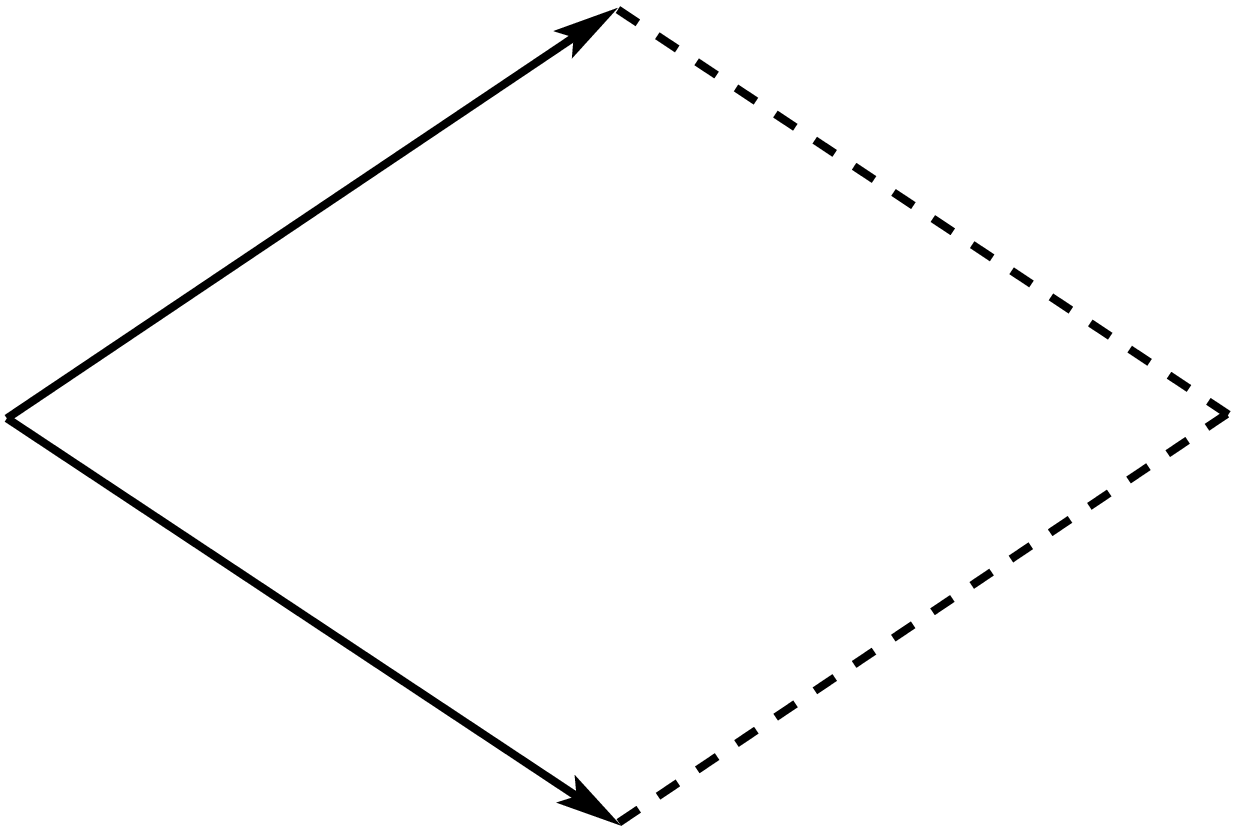}}

\mbox{\epsfxsize=6cm \epsffile{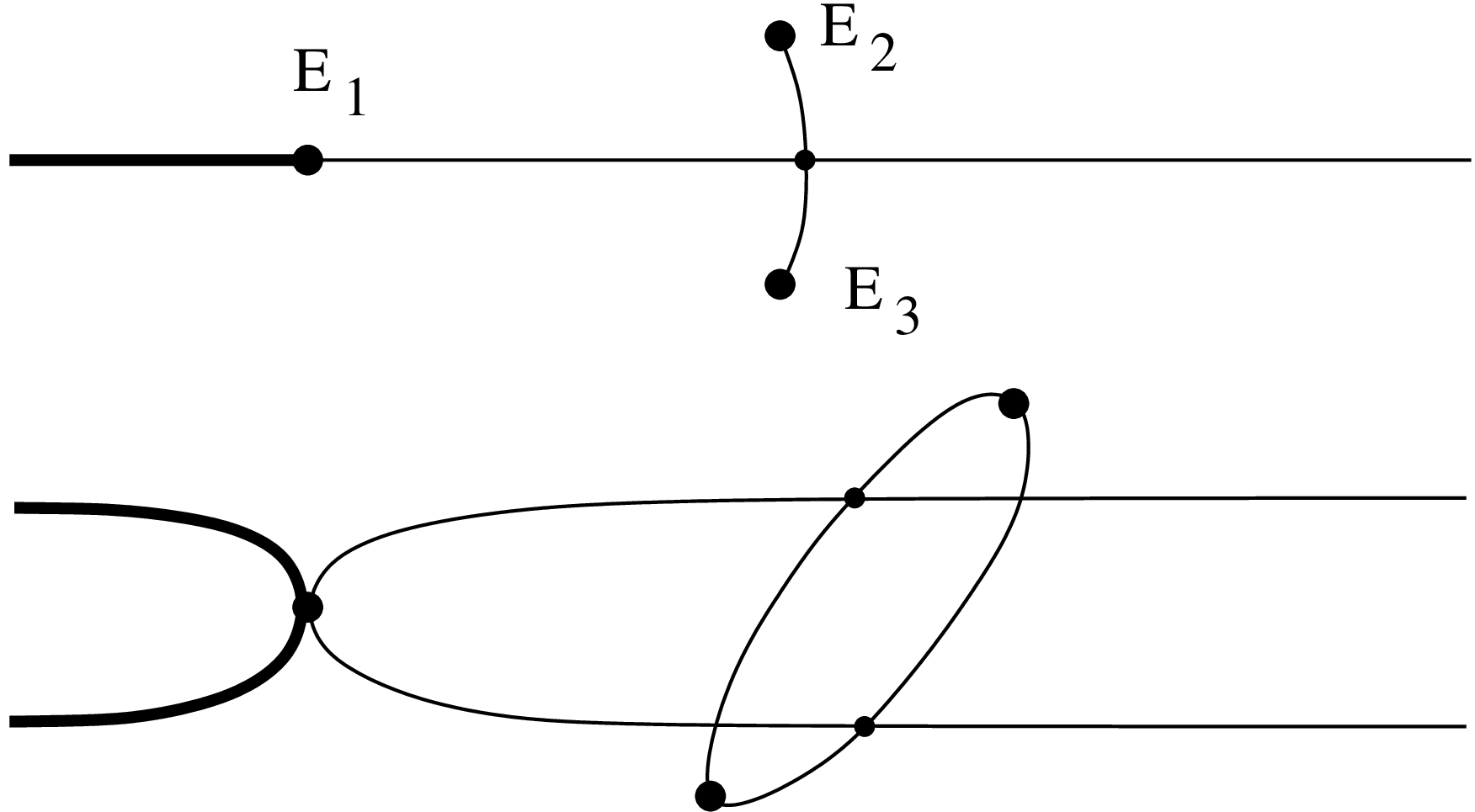}}

\parbox{6cm}{\begin{center} Fig 2a\\ The rhombic lattice \end{center}}
\parbox{6cm}{\begin{center} Fig 2b \\ The contour $|\varkappa|=1$  is singular \end{center}}
\end{center}

The classical Lam\'e problems do not lead to this case,  so it
never was considered.
\section*{Appendix~1. Proof of Theorem 1}
\addtocounter{section}{1}
\setcounter{subsection}{0}

 We prove Theorem 1 here and construct an  analog of the continuous
Fourier decomposition by the eigenfunctions of
 singular finite-gap operator with a
{\bf periodic} potential.

The proof consists of two steps.

1. We reduce  decomposition problem for the real singular potential to the
decomposition problem for regular complex potentials.

2. We construct eigenfunction expansion for regular complex potentials.

\subsection{Notations}

Let us recall some basic definitions.
 The spectral curve $\Gamma$ is defined by:
$$
\mu^2=(\lambda-E_0) \ldots (\lambda-E_{2g})=R(\lambda).
$$

Our divisor is $\cD=\gamma_1+\ldots+\gamma_g$.

The following  notations will be used in Appendix:
$\lambda(\gamma)$ denotes the projection of the
point $\gamma\in\Gamma$ to the $\lambda$-plane. So either
$\gamma=(\lambda(\gamma),+)$ or $\gamma=(\lambda(\gamma),-)$.

Let $\lambda_1=\lambda(\gamma_1)$,\ldots, $\lambda_g=\lambda(\gamma_g)$.

The quasimomentum differential $dp$ is uniquely defined by the
 following properties:
\begin{enumerate}
\item $dp$ is holomorphic in $\Gamma$ outside the point $\lambda=\infty$.
\item
$$
dp=dk\left(1+O\left(\frac{1}{k^2} \right) \right), \ \ k^2=\lambda
$$
near the point $\lambda=\infty$.
\item
Integrals over all basic cycles are purely real
\begin{equation}
\label{eq:normdp}
\Im \oint\limits_{c} dp =0
\end{equation}
for any closed contour $c\subset\Gamma$
\end{enumerate}
The quasimomentum function $p(\gamma)$ is the primitive of $dp$, and it is always
multivalued. We assume, that
$$
p(\gamma)=k+O\left(\frac{1}{k} \right).
$$

From (\ref{eq:normdp}) it follows, that the imaginary part of the
 quasimomentum function
$\Im p(\gamma)$ is well-defined.

We assume, that our potential $u(x)$ is periodic with the period $T$.
It implies, that
$$
\exp(ip(\gamma) T)
$$
is a single-valued function in $\Gamma$.

As above, we  denote the Bloch function by $\Psi(\gamma,x)$,
and $\sigma$ denotes the
holomorphic involution, interchanging the sheets of the surface $\Gamma$:
$$
\sigma: (\lambda,+)\rightarrow (\lambda,-), \ \ \Psi^*(\gamma,x)=
\Psi(\sigma\gamma,x)
$$

Assume, that function $f(x)$ has finite support and $f\in\cF_{X}$.

Let us define the continuous Fourier transform for $f(x)$ by:
\begin{equation}
\label{eq:contFourier}
\hat f(\gamma)= \frac{1}{2\pi} \int\limits_{-\infty}^{\infty} \Psi^*(y,\gamma) f(y) dy
\end{equation}
where we use the rule of going around the singularities within the complex domain.

{\bf Theorem 1''.}
{\it Assume, that:
\begin{enumerate}
\item The spectral curve $\Gamma$ is regular (has no multiple points),
\item The contour $\Im p(\gamma)=0$ is regular, i.e. $dp(\gamma)\ne0$ everywhere at this contour.
\end{enumerate}

Then we have the following reconstruction formula:

\begin{equation}
\label{eq:continv}
f(x) = \oint\limits_{\Im p(\gamma)=0} \hat f(\gamma) \Psi(x,\gamma)
\frac{(\lambda(\gamma)-\lambda_1)\ldots(\lambda(\gamma)-\lambda_g)}
{2\sqrt{(\lambda(\gamma)-E_0)\ldots(\lambda(\gamma)-E_{2g}))}}  d\lambda(\gamma).
\end{equation}

For any regular $x$ the integrand in (\ref{eq:continv}) decays for $\gamma\rightarrow\infty$ faster,
than any degree of $\lambda$.
}

\subsection{Reduction to
 smooth potential}

Apply now a series of Crum transformations. We intend to reduce
the decomposition with respect to singular real potential to the
decomposition with respect to
complex nonsingular potential.

\begin{lem}
Let $n=n_{\mbox{\scriptsize max}}$ denotes the maximal order of singularity
$n_{\mbox{\scriptsize max}}=\max\limits_j n_j$.  Then applying a series of $n$
properly chosen Darboux--Crum tranformation one can obtain a
 regular {\bf complex}
potential.
\end{lem}
{\bf Proof.}

Consider the image of all divisor trajectories in $\Gamma$. The
potential $u(x)$
is periodic. Therefore they form a compact set. Let $q_1\in\Gamma$
be a point outside of this set,
 such that $\Im p(q_1) \ne 0$, $l_1=\lambda(q_1)$.

Let $\psi_1(x)=\Psi(q_1,x)$ be the corresponding Baker--Akhiezer function.
$$
L\psi_1(x) = l_1 \psi_1(x).
$$
Then
\begin{enumerate}
\item
$
\psi_1(x) = \frac{1}{(x-x_j)^{n_j}}(a^{(j)}_0+o(1)), \ \ a^{(j)}_0 \ne 0
$
at all singular points.
\item $\psi_1(x)\ne0$ for all $x\in\RR$, $x\ne x_j$.
\end{enumerate}

Let
$$
Q_1=\left(\partial_x -\frac{\psi_x}{\psi} \right), \ \
Q_1^*=\left(-\partial_x -\frac{\psi_x}{\psi} \right).
$$
We have
$$
L-l_1= Q_1^* Q_1, \ \ L_1-l_1= Q_1 Q_1^*,
$$
where $L_1$ denotes the (Darboux-Crum)-transformed operator with potential
$$
u^{(1)}(x)=u(x)-2\partial^2_x \log(\psi_1)
$$
and Bloch function
$$
\Psi^{(1)}(x,\gamma) = \frac{1}{\lambda-l_1}Q_1\Psi(x,\gamma).
$$
We see, that this transformation reduces the orders of all
singularities by 1 and generates no new singular points.
By repeating this procedure $n$ times we come to the smooth potential
$u^{(n)}(x)$.

Let us define the  operators $L=L_0$, $L_1$, \ldots, $L_n$ by the following
formulas
$$
L_n=-\partial_x^2+u^{(n)}(x), \ \ L_k-l_{k}=Q_kQ_k^*, \ \ L_k-l_{k+1}= Q_{k+1}^* Q_{k+1},
$$
$$
Q_k Q_k^* = Q_{k+1}^* Q_{k+1}+l_{k+1}-l_k
$$

This procedure generates Bloch functions with
 slightly non-standard normalization.
To obtain the standard Baker--Akhiezer function, it is necessary
to change  normalization of
$\Psi(x,k)$.

Let us denote by $\gamma_1(x)$, \ldots, $\gamma_g(x)$ the divisor of
 zeroes
of $\Psi(x,\gamma)$.

Let $x_0$ be one of the singular points with the highest order singularity.
 It means that
for $x=x_0$ exactly $n$ points of the divisor $\gamma_1(x_0)$, \ldots,
$\gamma_g(x_0)$ are located
at the point $\lambda=\infty$. Denote the remaining points by
$\gamma_1(x_0)$, \ldots,
$\gamma_{g-n}(x_0)$.

Let $\tilde\Psi(x,\gamma)$ be the Baker--Akhiezer function with $g-n$
 simple poles $\gamma_1(x_0)$, \ldots,
$\gamma_{g-n}(x_0)$ at the finite part of $\Gamma$. It has the asymptotics
$\tilde\Psi(x,\gamma) = e^{ik(x-x_0)}((-ik)^n+O(k^{n-1})), \ \
k^2=\lambda, \ \ \lambda\rightarrow\infty.$

Then
\begin{equation}
\tilde\Psi^{(n)}(x,\gamma) =\frac{1}{(\lambda-l_1)\cdots(\lambda-l_n)}\cdot
Q_nQ_{n-1}\ldots Q_1 \tilde\Psi(x,\gamma)
\end{equation}
is the Baker--Akhiezer function for the smooth operator $L_n$ with
 the divisor of poles
$\gamma_1(x_0)$, \ldots, $\gamma_{g-n}(x_0)$, $\sigma q_1$, \ldots,
$\sigma q_n$ and essential singularity
\begin{equation}
\tilde\Psi^{(n)}(x,\gamma) = e^{ik(x-x_0)}(1+o(1)), \ \
 \lambda\rightarrow\infty.
\end{equation}

\begin{lem}
The operators $Q_j$, $Q_J^*$ map Bloch functions to the Bloch
functions with the same multiplier.
\end{lem}

Consider the following operator:
$$
M=Q_n\cdot\ldots\cdot Q_1\cdot Q^*_1\cdot\ldots\cdot Q^*_n
$$
\begin{lem}
We have the following formula
$$
M=(L_n-l_1)(L_n-l_2)\ldots(L_n-l_n)
$$
\end{lem}
The proof is straightforward:
$$
Q_n\ldots Q_4 Q_3 Q_2 Q_1 Q_1^*Q_2^*Q_3^*Q_4^*\ldots Q_n^*
=Q_n\ldots Q_4 Q_3 Q_2 (Q_2^* Q_2+l_2-l_1) Q_2^*Q_3^*Q_4^*\ldots Q_n^* =
$$
$$
=Q_n\ldots Q_4 Q_3 (Q_2 Q_2^*+l_2-l_1) Q_2Q_2^*Q_3^*Q_4^*\ldots Q_n^* =
$$
$$
= Q_n\ldots Q_4 Q_3 (Q_3^* Q_3+l_3-l_1) (Q_3^*Q_3+l_3-l_2) Q_3^*Q_4^* \ldots Q_n^* =
$$
$$
= Q_n\ldots Q_4 (Q_3 Q_3^* +l_3-l_1) (Q_3Q_3^*+l_3-l_2) Q_3 Q_3^*Q_4^*\ldots Q_n^*  =
$$
$$
= Q_n\ldots Q_4(Q_4^* Q_4 +l_4-l_1) (Q_4^*Q_4+l_4-l_2) (Q_4^* Q_4+l_4-l_3) Q_4^* \ldots Q_n^* =
$$
$$
= Q_n\ldots (Q_4 Q_4^* +l_4-l_1) (Q_4Q_4^*+l_4-l_2) (Q_4 Q_4^*+l_4-l_3) Q_4Q_4^* \ldots Q_n^* =
$$
$$
\ldots
$$
$$
=(Q_n Q_n^* +l_n-l_1) (Q_nQ_n^*+l_n-l_2)\ldots  (Q_n Q_n^*+l_n-l_{n-1}) Q_nQ_n^*=
$$
$$
=(L_n-l_1)(L_n-l_2)\ldots (L_n-l_{n-1})(L_n-l_n)
$$
\begin{cor}
The operator $M$ is a differential operator with smooth coefficients.
\end{cor}

\begin{rem}
It follows from the definition of $Q_k$  that
\begin{equation}
Q^*_1\cdot\ldots\cdot Q^*_n\cdot Q_n\cdot\ldots\cdot Q_1 \Psi(x,\gamma)=
(\lambda(\gamma)-l_n)\ldots(\lambda(\gamma)-l_1)\Psi(x,\gamma)
\end{equation}
\end{rem}

\begin{lem}
Let:
$$
f^{(n)}(x) = Q_n\cdot Q_{n-1}\cdot\ldots\cdot Q_1 f(x),
$$
where $f\in\cF_{X}$. Then $f^{(n)}(x)$ is a complex smooth
 periodic function.

\end{lem}

The proof is straightforward: Each operator $Q_k$ reduces
the order $n_j$ of singularity at the point
$x_j$ by 1.

\begin{lem}
Assume, that the function  $f^{(n)}(x)$ admits the Laurent-Fourier
decomposition
in the  Bloch functions for the operator $L_n$:
$$
f^{(n)}(x)= \sum\limits_j c_j \tilde\Psi^{(n)}(\kappa_j,x),
$$
Then the function   $M^{-1}f^{(n)}(x)$ is well-defined, and we have
$$
M^{-1}f^{(n)}(x) =  \sum\limits_j \frac{c_j}
{(\lambda(\kappa_j)-l_1)\ldots(\lambda(\kappa_j)-l_n))   }
\tilde\Psi^{(n)}(\kappa_j,x).
$$
For all $\kappa_j$ we have $\Im p(\kappa_j)=0$.
Therefore we have no zeroes in the denominators.
\end{lem}
We also have:
$$
f^{(n)}(x) = M(M^{-1} f^{(n)}(x)) =
Q_n\cdot\ldots\cdot Q_1\cdot Q^*_1\cdot\ldots\cdot Q^*_n\cdot (M^{-1}f^{(n)}(x)),
$$
and therefore
\begin{equation}
\label{eq:decomp1}
f(x)= Q^*_1\cdot\ldots\cdot Q^*_n\cdot (M^{-1}f^{(n)}(x)).
\end{equation}

The results of this section can be summarized in the following way.
 To decompose any given function
$f(x)$ we perform following actions:

\begin{enumerate}
\item By applying $n$ properly chosen Darboux-Crum transformations
 we obtain a smooth functions
 $f^{(n)}(x) = Q_n\cdot Q_{n-1}\cdot\ldots\cdot Q_1 f(x)$.
\item We decompose the smooth function $f^{(n)}(x)$ in the  eigenfunctions
of {\bf the smooth complex
finite-gap} operator $L_n$.
\item Taking into account, that all points $q_n$ are located outside of
the contour $\Im(p)=0$,
we construct the eigenfunctions decomposition for $M^{-1}f_n(x)$.
\item Applying formula (\ref{eq:decomp1}) we obtain an eigenfunction
decomposition of the original function $f(x)$. At this
stage we use the following property of the Darboux-Crum transformation:
the operator
$Q^*_1\cdot\ldots\cdot Q^*_n\cdot (M^{-1}f^{(n)}(x))$ maps the Bloch eigenfunctions of $L_n$ to the
eigenfunctions of $L$.
\end{enumerate}

To complete the proof, it is sufficient to prove decompositions theorem for smooth {\bf complex}
finite-gap operators.

\subsection{Decomposition for smooth complex periodic potentials}

{\bf Remark.} Recently I.M. Krichever pointed out to the authors that the 
proof of eigenfunctions decomposition  theorem for {\bf regular complex} 
finite-gap potentials was proved in his paper \cite{Krich_FA_1986}.

Consider a hyperelliptic Riemann surface $\Gamma$ with divisor $\cD$.
We assume, that the
corresponding potential $u(x)$ is regular and periodic with the period $T$, but may be complex.

 It is natural to use the local coordinate $1/k$ near infinity  where
 $k=p(\gamma)$,
$\lambda=k^2+O(1)$. In the neighbourhood of infinity we have:
$$
\Psi(x,\gamma) = e^{ikx}\left[1+\frac{\phi_1(x)}{k}+O\left(\frac{1}{k^2} \right) \right], \ \
\Psi^*(y,\gamma) = e^{-iky}\left[1-\frac{\phi_1(y)}{k}+O\left(\frac{1}{k^2} \right) \right], \ \
$$
$$
\phi_1(x+T) = \phi_1(x).
$$
Let us denote
$$
\Xi(x,y,\gamma) = \Psi(x,\gamma) \Psi^*(y,\gamma)
\frac{(\lambda(\gamma)-\lambda_1)\ldots(\lambda(\gamma)-\lambda_g)}
{2\sqrt{(\lambda(\gamma)-E_0)\ldots(\lambda(\gamma)-E_{2g}))}}  d\lambda(\gamma)
$$
There exists a constant $K_0$ such that

 (1) In the domain $|k|>K_0$
the function $1/k$ is a well-defined
local coordinate.

(2) The function $\Psi(x,\gamma) e^{-ikx} $
is holomorphic in the variable $1/k$ for $x$ in the domain
$|k|>K_0$, $|\Im x|<\epsilon$.

For sufficiently large $k$ we have
$$
\Xi(x,y,\gamma) = e^{ik(x-y)}\left[1+\frac{\phi_1(x)-
\phi_1(y)}{k}+\frac{\chi_2(k,x,y)}{k^2}\right] dk
$$
where $\chi_2(x,y,k)$ is holomorphic in $1/k$, $x$, $y$
in the domain $|k|>K_0$, $|\Im x|<\epsilon$,
$|\Im y|<\epsilon$.

Our purpose is to study the convergence of the Fourier transformation.
 Let $f(x)$ be either a
Schwartz class function or a Bloch-periodic function.

\begin{enumerate}
\item Case 1. The integral Fourier transform.
Let $f(x)$ be a function with a finite support.

We define:
$$
\hat f(\gamma)= \frac{1}{2\pi} \int\limits_{-\infty}^{\infty} \Psi^*(y,\gamma) f(y) dy
$$
\begin{equation}
\label{eq:contS}
(\hat\cS(K)f)(x) = \oint\limits_{\begin{array}{c}\scriptstyle\Im p(\gamma)=0, \\
\scriptstyle\ |\Re p(\gamma)|\le K \end{array}} \hat f(\gamma) \Psi(x,\gamma)
\frac{(\lambda(\gamma)-\lambda_1)\ldots(\lambda(\gamma)-\lambda_g)}
{2\sqrt{(\lambda(\gamma)-E_0)\ldots(\lambda(\gamma)-E_{2g}))}}  d\lambda(\gamma)
\end{equation}
Our purpose is to show, that $(\hat\cS(K)f)(x)$ converges to $f(x)$
as $K\rightarrow\infty$.
It is easy to to see that
$$
(\hat\cS(K)f)(x) = \frac{1}{2\pi} \int\limits_{-\infty}^{\infty} S(K,x,y)
 f(y) dy
$$
where
\begin{equation}
\label{eq:dacayingS}
S(K;x,y)=\oint\limits_{\begin{array}{c}\scriptstyle\Im p(\gamma)=0, \\
\scriptstyle\ |\Re p(\gamma)|\le K \end{array}} \Xi(x,y,\gamma)
\end{equation}
In this section we shall prove the following theorem:
\begin{prop}
\label{thm:contKer}
Let us assume that:
\begin{enumerate}
\item The spectral curve $\Gamma$ is regular (i.e. has no multiple points),
\item The contour $\Im p(\gamma)=0$ is regular, i.e. $dp(\gamma)\ne0$
everywhere at this contour.
\end{enumerate}
Then:
\begin{enumerate}
\item The kernel $S(K,x,y)$ has the following structure
$$
S(K,x,y)=S_{\mbox{classical}}(K,x,y)+S_{\mbox{correction}}(K,x,y)
$$
where
$$
S_{\mbox{classical}}(K,x,y)=\frac{2\sin(K(x-y))}{x-y}
$$
is the corresponding kernel for the ``standard'' integral Fourier transform, and
$S_{\mbox{correction}}(K,x,y)$ uniformly converges at any compact set in the
$(x,y)$-plane to a continuous function $S_{\mbox{correction}}(\infty,x,y)$.
\item Let $x$ does not belong to the support of $f(y)$. Then $(\hat\cS(K)f)(x)\rightarrow 0$ for
$K\rightarrow\infty$, and $S_{\mbox{correction}}(\infty,x,y)\equiv0$. Moreover
$(\hat\cS(K)f)(x)\rightarrow 0$ faster than any degree of $K$.
\end{enumerate}
\end{prop}
\item Case 2. The discrete Fourier transform.
 Let $f(x)$ be Bloch-periodic with the period $T$:
$$
f(x+T)=\varkappa_0 f(x),
$$
where $\varkappa_0=e^{iT\varphi_0}$ is an unitary multiplier  $|\varkappa_0|=1$. Consider the set of all
points  $\kappa_j$ such, that $e^{iTp(\kappa_j)}=\varkappa_0$.
Let us define
$$
\hat f(\kappa_j)= \frac{1}{T} \int\limits_{0}^{T} \Psi^*(\kappa_j,y) f(y) dy
$$
The multipliers in the integrand have opposite Bloch multipliers, therefore we can integrate over
any basic period.
Let us define
$$
(\hat\cS(N)f)(x) = \!\!\!\!\!\!\!\!\!\! \!\!\!\! \sum\limits_{|(p(\kappa_j)-\varphi_0)T|\le 2\pi N}
 \!\!\!\!\!\!\!\!\!\! \!\!\!\! \hat f(\kappa_j) \Psi(\kappa_j,x)
\frac{(\lambda(\kappa_j)-\lambda_1)\ldots(\lambda(\kappa_j)-\lambda_g)}
{2\sqrt{(\lambda(\kappa_j)-E_0)\ldots(\lambda(\kappa_j)-E_{2g}))}}
\left.\left[\frac{  d\lambda(\gamma)}{dp(\gamma)}
\right]\right|_{\lambda=\kappa_j}
$$
We have
$$
(\hat\cS(N)f)(x) = \int\limits_{0}^{T} S(N,x,y) f(y) dy
$$
where
\begin{equation}
\label{eq:blochS}
S(N,x,y)=\frac{1}{T}\sum\limits_{|(p(\kappa_j)-\varphi_0)T|\le 2\pi N} \frac{\Xi(\kappa_j,x,y)}{dp(\kappa_j)}
\end{equation}
\begin{prop}
\label{thm:discrKer}
Assume, that all points $\kappa_j$ such that
$e^{iTp(\kappa_j)}=\varkappa_0$,are regular:
\begin{enumerate}
\item They do not coincide with the multiple points (if they exists).
\item $dp(\kappa_j)\ne0$ for all $j$.
\end{enumerate}
Then
\begin{enumerate}
\item The kernel $S(N,x,y)$ has the following structure
$$
S(N,x,y)=S_{\mbox{classical}}(N,x,y)+S_{\mbox{correction}}(N,x,y)
$$
where
$$
S_{\mbox{classical}}(N,x,y)=\frac{e^{i\phi_0(x-y)}}{T} \frac{\sin{\left(\frac{\pi(2N+1)}{T}(x-y) \right)}}
{\sin{\left(\frac{\pi}{T}(x-y) \right)}}
$$
is the corresponding kernel for the ``standard'' discrete
 Fourier transform and
$S_{\mbox{correction}}(N,x,y)$ uniformly converges in the $(x,y)$-plane
 to the continuous function
$S_{\mbox{correction}}(\infty,x,y)$.
\item Let a point $x$ does not belong to the support of $f(y)$.
Then $(\hat\cS(N)f)(x)\rightarrow 0$ for
$N\rightarrow\infty$, and $S_{\mbox{correction}}(\infty,x,y)\equiv0$.
\end{enumerate}
\end{prop}
\end{enumerate}

We prove now the first part of Proposition~\ref{thm:contKer}.

Let $S(K,x,y)$ be the kernel defined by formula (\ref{eq:dacayingS})

We assume, that the orientation on this contour is defined by \mbox{$\Re dp(\gamma)>0$}.
Let us fix a sufficiently large constant $K_0$. Then we can write
$$
S(K;x,y)=I_1(x,y)+I_2(K,x,y)+I_3(K,x,y)+I_4(K,x,y)
$$
$$
I_1(x,y)=\oint\limits_{\begin{array}{c}\scriptstyle\Im p(\gamma)=0, \\
\scriptstyle|\Re p(\gamma)|\le K_0 \end{array}} \Xi(x,y,\gamma)
$$
$$
I_2(K,x,y)=\left[\int\limits_{-K}^{-K_0}+\int\limits_{K_0}^{K}\right]
e^{ik(x-y)} dk
$$
$$
I_3(K,x,y)=\left[\int\limits_{-K}^{-K_0}+\int\limits_{K_0}^{K}\right]
e^{ik(x-y)}\left[\frac{\phi_1(x)-\phi_1(y)}{k} \right] dk
$$
$$
I_4(K,x,y)=\left[\int\limits_{-K}^{-K_0}+\int\limits_{K_0}^{K}\right]
e^{ik(x-y)} \frac{\chi_2(k,x,y)}{k^2}  dk
$$
A standard calculation implies:
$$
I_2(K,x,y)=\frac{2\sin(K(x-y))}{x-y}-\frac{2\sin(K_0(x-y))}{x-y}
$$
Let us denote:
$$
S_{\mbox{classical}}(K,x,y)=I_2(K,x,y)+\frac{2\sin(K_0(x-y))}{x-y}
$$
$$
S_{\mbox{correction}}(K,x,y)=I_1(x,y)+I_3(K,x,y)+I_4(K,x,y)-\frac{2\sin(K_0(x-y))}{x-y}
$$

The functions $I_1(x,y)$ and $-\frac{2\sin(K_0(x-y))}{x-y}$ do not depend
 on $K$ and
are continuous in both variables. Integral $I_4(K,x,y)$
absolutely converges as $K\rightarrow\infty$, therefore the
limiting function is
continuous in $x$, $y$. We also have
$$
I_3(K,x,y)=\frac{2}{i}{\rm Si\,}(K(x-y))(\phi_1(x)-\phi_1(y))  -
\frac{2}{i}{\rm Si\,}(K_0(x-y))(\phi_1(x)-\phi_1(y)),
$$
where
$$
{\rm Si\,}(x) = \int\limits_{0}^{x}\frac{\sin(t)}{t} dt,
$$
therefore it uniformly converges to a continuous function
$$
I_3(\infty,x,y)=\frac{\pi}{i} \mbox{sign}(x-y))(\phi_1(x)-\phi_1(y))  -
\frac{2}{i}{\rm Si\,}(K_0(x-y))(\phi_1(x)-\phi_1(y)),
$$
This completes the proof. \qed

\begin{cor}
The kernel $S(\infty,x,y)=\lim\limits_{K\rightarrow\infty}S(K,x,y)$
is a well-defined
distribution and
$$
S(\infty,x,y)=2\pi\delta(x-y)+ S_{\mbox{correction}}(\infty,x,y)
$$
\end{cor}

Proof of  the first part of Proposition~\ref{thm:discrKer}.

Consider a sufficiently large $N_0$. It is natural to write
$$
S(N,x,y)=I_1(x,y)+I_2(N,x,y)+I_3(N,x,y)+I_4(N,x,y)
$$
where
$$
I_1(x,y)=\frac{1}{T}\sum\limits_{|(p(\kappa_j)-\varphi_0)T|\le 2\pi N_0}
 \frac{\Xi(\kappa_j,x,y)}{dp(\kappa_j)},
$$
$$
I_2(N,x,y)=\frac{1}{T}\left[ \sum\limits_{j=-N}^{-1-N_0}+\sum\limits_{j=N_0+1}^N \right]
e^{\left(\frac{2\pi i}{T}N+i\varphi_0\right) (x-y)}
$$
$$
I_3(N,x,y)=\frac{1}{T}\left[ \sum\limits_{j=-N}^{-1-N_0}+\sum\limits_{j=N_0+1}^N \right]
\frac{e^{\left(\frac{2\pi i}{T}j+i\varphi_0\right) (x-y)}}{\frac{2\pi}{T}j+\varphi_0}(\phi_1(x)-\phi_1(y))
$$
$$
I_4(N,x,y)=\frac{1}{T}\left[ \sum\limits_{j=-N}^{-1-N_0}+\sum\limits_{j=N_0+1}^N \right]
e^{\left(\frac{2\pi i}{T}j+i\varphi_0\right) (x-y)}
\frac{\chi_2({2\pi}{T}j+\varphi_0,x,y)}{(\frac{2\pi}{T}j+\varphi_0)^2}
$$
A standard calculation implies:
$$
I_2(N,x,y)=\frac{e^{i\varphi_0(x-y)}}{T} \frac{\sin{\left(\frac{\pi(2N+1)x}{T} \right)}}
{\sin{\left(\frac{\pi x}{T} \right)}} -
\frac{e^{i\varphi_0(x-y)}}{T} \frac{\sin{\left(\frac{\pi(2N_0+1)x}{T} \right)}}
{\sin{\left(\frac{\pi x}{T} \right)}}
$$
Let us denote:
$$
S_{\mbox{classical}}(N,x,y)=I_2(N,x,y)+ \frac{e^{i\varphi_0(x-y)}}{T}
\frac{\sin{\left(\frac{\pi(2N_0+1)x}{T} \right)}}
{\sin{\left(\frac{\pi x}{T} \right)}}=
\frac{e^{i\varphi_0(x-y)}}{T} \frac{\sin{\left(\frac{\pi(2N+1)x}{T} \right)}}
{\sin{\left(\frac{\pi x}{T} \right)}}
$$
$$
S_{\mbox{correction}}(N,x,y)=I_1(x,y)+I_3(N,x,y)+I_4(N,x,y)-\frac{e^{i\varphi_0(x-y)}}{T}
\frac{\sin{\left(\frac{\pi(2N_0+1)x}{T} \right)}}{\sin{\left(\frac{\pi x}{T} \right)}}
$$

The term $I_1(x,y)$ is continuous in $x$, $y$, $I_4(N,x,y)$ uniformly
 converges to a
continuous function.

The term  $I_3(N,x,y)$ requires some extra attention. It can be written as:
$$
I_3(N,x,y)=\frac{1}{2\pi}e^{i\varphi_0(x-y)}\left[ \sum\limits_{j=-N}^{1-N_0}+\sum\limits_{j=N_0+1}^N \right]
\frac{e^{\left(\frac{2\pi i}{T}j\right) (x-y)}}{j}(\phi_1(x)-\phi_1(y))+
$$
$$
-\frac{\varphi_0}{T}e^{i\varphi_0(x-y)}\left[ \sum\limits_{j=-N}^{1-N_0}
+\sum\limits_{j=N_0+1}^N \right]
\frac{e^{\left(\frac{2\pi i}{T}j\right) (x-y)}}{(\frac{2\pi}{T}j+\varphi_0)
(\frac{2\pi}{T}j)}(\phi_1(x)-\phi_1(y))
$$
The second term uniformly converges to a continuous function in $x$, $y$.

Let us denote
$$
S_1(N,z) = \left[ \sum\limits_{k=-N}^{1}+\sum\limits_{k=1}^N \right]
\frac{e^{ikz}}{k} =
i \int\limits_{0}^{z} \left[\frac{\sin{\left(\left( N+ \frac{1}{2}\right)
 w  \right)}}
{\sin{\left(\frac{w}{2} \right)}} -1\right] dw
$$
Function $S_1(N,z)$ is periodic with period $2\pi$ and converges to
$i(\pi\mbox{sign}(z)-z)$ at the
interval $[-\pi,\pi]$ uniformly outside any neighborhood of the point
$z=0$. We have
$$
I_3(N,x,y)=\frac{1}{2\pi}e^{i\varphi_0(x-y)}S_1(N,\frac{2\pi}{T}(x-y))
(\phi_1(x)-\phi_1(y))+
\mbox{regular terms}
$$
therefore it also converges uniformly to a continuous kernel. \qed
\begin{cor}
The kernel $S(\infty,x,y)=\lim\limits_{N\rightarrow\infty}S(N,x,y)$ is
a well-defined
distribution and
$$
S(\infty,x,y)=\sum\limits_j\delta(x-y-jT)+ S_{\mbox{correction}}(\infty,x,y)
$$
\end{cor}

To continue the proof we need the following:
\begin{lem}
\label{lem:est1}
Let $f(y)$ be a smooth function with compact support.
$$
\hat f(\gamma)= \frac{1}{2\pi} \int\limits_{-\infty}^{\infty} \Psi^*(y,
\gamma) f(y) dy
$$
Then for any $n$ there exists a constant $C_n=C_n(u[y],f[y])$ such,
 that for
sufficiently large $\lambda(\gamma)$ we have
$$
|\hat f(\gamma)| \le \frac{C_n}{|\lambda(\gamma)|^n} \ \
 \max\limits_{y\in\mbox{\scriptsize supp}\
f(y)} \ \  |e^{-ip(\gamma)y}|
$$
\end{lem}
{\bf Proof.} By definition, we have $L^n \Psi^*(x,\gamma)=\lambda(\gamma)^n\Psi^*(x,\gamma)$. Therefore
$$
\lambda(\gamma)^n \int\limits_{-\infty}^{\infty} \Psi^*(y,\gamma) f(y) dy =
\int\limits_{-\infty}^{\infty}[L^n \Psi^*(y,\gamma)] f(y) dy.
$$
Function $f(y)$ has a finite support, therefore we can eliminate
 all derivatives of
$\Psi^*(y,\gamma)$ by integrating by parts, and we obtain
$$
\lambda(\gamma)^n \int\limits_{-\infty}^{\infty} \Psi^*(y,\gamma) f(y)
dy =
$$
$$
=\int\limits_{-\infty}^{\infty} \Psi^*(y,\gamma)
P_n(f(y),f'(y),\ldots,f^{(2n)}(y),u(y),u'(y),\ldots,u^{(2n-2)}(y))   dy.
$$
where $P(\ldots)$ is a polynomial. The function $e^{ip(\gamma)y}
\Psi^*(y,\gamma)$
is uniformly bounded for all $y$ and sufficiently large $\lambda(\gamma)$,
therefore we obtain the desired estimate.

Let us prove the second part of Proposition~\ref{thm:contKer}.

Consider the integral representation  (\ref{eq:contS})
for $(\hat\cS(K)f)(x)$.
The integrand is holomorphic in $\gamma$ at the finite part of $\Gamma$.
From Lemma~\ref{lem:est1} we see that the integral (\ref{eq:contS})
 absolutely
converges as $K\rightarrow\infty$ in the upper half-plane if
$x>\mbox{supp}\ f(y)$ or in the lower half-plane if $x<\mbox{supp}\ f(y)$.
Moreover the integrand exponentially decreases in the corresponding
half-plane,
therefore it is equal to 0. The integrand decays at infinity faster
 then any
degree of $K$, therefore the integral is fast decaying as
$K\rightarrow\infty$.

In order to finish the proof of the second part of
 Proposition~\ref{thm:discrKer}, we shall use the
following integral representation for $S(N,x,y)$
$$
S(N,x,y)=\frac{1}{2\pi}\oint\limits_{\beta_N}
\frac{\Xi(\kappa_j,x,y)}{e^{ip(\kappa_j)T}-\varkappa_0}
$$
Here $\beta_N$ denotes the following contour
 (we assume $N$ to be sufficiently large
and $p(\gamma)$ is fixed near infinity as a single-valued function):
$$
\beta_N=\beta_N^{(1)}\cup \beta_N^{(2)}\cup \beta_N^{(3)}\cup \beta_N^{(4)}
$$
$$
\beta_N^{(1)}=\{\Im p(\gamma) =-N, |(\Re p(\kappa_j)-\varphi_0)T|\le 2\pi (N+1/2) \},
$$
$$
\beta_N^{(2)}=\{|\Im p(\gamma)| \le N, (\Re p(\kappa_j)-\varphi_0)T = 2\pi (N+1/2) \},
$$
$$
\beta_N^{(3)}=\{\Im p(\gamma) =N, |(\Re p(\kappa_j)-\varphi_0)T|\le 2\pi N \},
$$
$$
\beta_N^{(4)}=\{|\Im p(\gamma)| \le N, (\Re p(\kappa_j)-\varphi_0)T = - 2\pi (N+1/2) \},
$$
We choose the orientation on $\beta_N$  by assuming, that infinite point
is located outside of
the contour.

By calculating the residues we immediately obtain formula
(\ref{eq:blochS}).

\begin{rem} At all multiple points (if they exist)
we have $\Im p(E_j) =0$.
Therefore all of them are inside the contour $\beta_N$.
 For any holomorphic
differential on  singular curve, the sum of residues at singular points
 is equal
to zero. Therefore they do not affect our integral.
\end{rem}

 We have

$$
(\hat\cS(N)f)(x) =  \frac{1}{2\pi}\oint\limits_{\beta_N} \int\limits_{x_0}^{x_0+T}
\frac{\Xi(\kappa_j,x,y)}{e^{ip(\kappa_j)T}-\varkappa_0} f(y) dy
$$

Let us choose $x=x_0$. The support of $f(y)$ does not contain $x$,
therefore we can write
$$
(\hat\cS(N)f)(x) =  \frac{1}{2\pi}\oint\limits_{\beta_N}
\int\limits_{x+\varepsilon}^{x+T-\varepsilon}
\frac{\Xi(\kappa_j,x,y)}{e^{ip(\kappa_j)T}-\varkappa_0} f(y) dy
$$
for some $\varepsilon>0$.

From Lemma~\ref{lem:est1} it follows, that for any $M$ there exist
 constants $D_M$
such, that
$$
\left|\int\limits_{x+\varepsilon}^{x+T-\varepsilon}
\frac{\Xi(\kappa_j,x,y)}{e^{ip(\kappa_j)T}-\varkappa_0} f(y) dy\right| \le D_M
e^{-N\epsilon T} \ \ \mbox{on} \ \ \beta_1, \beta_3
$$
$$
\left|\int\limits_{x+\varepsilon}^{x+T-\varepsilon}
\frac{\Xi(\kappa_j,x,y)}{e^{ip(\kappa_j)T}-\varkappa_0} f(y) dy\right| \le \frac{D_M}{N^M}
\ \ \mbox{on} \ \ \beta_2, \beta_4
$$
Therefore $(\hat\cS(N)f)(x)$ tends to $0$ faster than any degree
of $N$ as $N\rightarrow\infty$.

\subsection{The reconstruction formula for singular potentials}
To complete the proof, let us check formula (\ref{eq:continv}).
Let us denote
$$
\eta(\gamma)= \frac{\tilde\Psi(x,\gamma)}{\Psi(x,\gamma)}.
$$
Then
$$
\hat f(\gamma)= \frac{1}{2\pi \eta(\sigma\gamma)}
\int\limits_{-\infty}^{\infty} \tilde\Psi(y,\sigma\gamma) f(y) dy=
$$
$$
= \frac{1}{2\pi \eta(\sigma\gamma)}
\int\limits_{-\infty}^{\infty}
\frac{(Q^*_1\cdot\ldots\cdot Q^*_n\cdot Q_n\cdot\ldots\cdot Q_1\cdot
\tilde\Psi(y,\sigma\gamma)) f(y)}
{(\lambda(\gamma)-l_1)\ldots(\lambda(\gamma)-l_n)}dy=
$$
$$
= \frac{1}{2\pi \eta(\sigma\gamma)}
\int\limits_{-\infty}^{\infty}
\frac{( Q_n\cdot\ldots\cdot Q_1\cdot  \tilde\Psi(y,\sigma\gamma))
(Q_n\cdot\ldots\cdot Q_1 \cdot f(y))}
{(\lambda(\gamma)-l_1)\ldots(\lambda(\gamma)-l_n)}dy=
$$
$$
=\frac{1}{2\pi \eta(\sigma\gamma)} \int\limits_{-\infty}^{\infty}
\tilde\Psi^{(n)}(y,\sigma\gamma) f^{(n)}(y) dy=
\frac{1}{\eta(\sigma\gamma)} \hat f^{(n)}(\gamma).
$$

We have
$$
f^{(n)}(x)= \oint\limits_{\scriptstyle\Im p(\gamma)=0}
\hat f^{(n)}(\gamma) \tilde\Psi^{(n)}(x,\gamma)
\frac{(\lambda-\lambda_1(x_0))\ldots(\lambda-\lambda_{g-n}(x_0))
(\lambda-l_1)\ldots(\lambda-l_n)  }
{2\sqrt{(\lambda-E_0)\ldots(\lambda-E_{2g}))}}  d\lambda,
$$
where $\lambda=\lambda(\gamma)$.
$$
f(x)= Q^*_1\cdot\ldots\cdot Q^*_n\cdot (M^{-1}f^{(n)}(x)) =
$$
$$
=\oint\limits_{\scriptstyle\Im p(\gamma)=0}
\hat f^{(n)}(\gamma) \tilde\Psi(x,\gamma)
\frac{(\lambda-\lambda_1(x_0))\ldots(\lambda-\lambda_{g-n}(x_0))}
{2\sqrt{(\lambda-E_0)\ldots(\lambda-E_{2g}))}}  d\lambda=
$$
$$
=\oint\limits_{\scriptstyle\Im p(\gamma)=0}
\eta(\sigma\gamma)\eta(\gamma) \hat f(\gamma) \Psi(x,\gamma)
\frac{(\lambda-\lambda_1(x_0))\ldots(\lambda-\lambda_{g-n}(x_0))}
{2\sqrt{(\lambda-E_0)\ldots(\lambda-E_{2g}))}}  d\lambda=
$$
Taking into account, that
$$
\eta(\sigma\gamma)\eta(\gamma)=
\frac{(\lambda(\gamma)-\lambda_1)\ldots(\lambda(\gamma)-\lambda_g)}
{(\lambda-\lambda_1(x_0))\ldots(\lambda-\lambda_{g-n}(x_0))},
$$
we obtain the formula (\ref{eq:continv}).

\section*{Appendix~2. Proof of Statement~\ref{stat:st3} }
\addtocounter{section}{1}
\setcounter{subsection}{0}

In this Appendix we present a proof of the Statement~\ref{stat:st3}.

To be precise, we prove the following:

\begin{thm}
Assume, that we have the space $\cF_X$ associated with either a
decaying at infinity  potential with $N$ singular points of
orders $n_1$,\ldots,$n_N$ or a periodic potentials with $N$ singular points
of orders $n_1$,\ldots,$n_N$ at the period. In the periodic case we assume
that an unitary Bloch multiplier $\varkappa_0$, $|\varkappa_0|=1$ is fixed.

\begin{enumerate}
\item Denote by $l_X$ the following number  $l_X=\left[\frac{n_1+1}{2}\right]+\left[\frac{n_2+1}{2}\right]+\ldots+\left[\frac{n_N+1}{2} \right]$, where $[\ \ ]$
is the integer part of a number.

There exists an $l_X$-dimensional subspace of $\cF_X$
such that our scalar product is negative defined on it.
\item Any subspace of dimension $d>l_x$ has non-zero intersection with
$\cF_X^0$, i.e. contains at least one function with positive square.
\end{enumerate}
\end{thm}
The proof of the second part is straightforward. A function from the space
$\cF_X$ lies in $\cF_X^0$ if it satisfies exactly $l_x$ linear equations: all
singular terms in the expansions near points $x_j$ are equal to 0. We have
$d$-dimensional subspace with  $d>l_X$, therefore this system of equations
has a least one non-trivial solution.

To prove the first part of the Theorem we construct these negative
subspaces explicitly. In is convenient to consider the decaying and the
periodic cases separately.

\subsection{Decaying at infinity case.}

Let us prove three technical lemmas.

\begin{lem}
\label{lem:l7}
Assume, that we have only one singular point $x_1=0$ of
order $n$. For any $n$ the functions $1/x^{n-2l}$,
$l=0,1,\ldots,\left[\frac{n-1}{2}\right]$  generate a zero subspace with respect to
our scalar product:
$$
<\frac{1}{x^{n-2k}},\frac{1}{x^{n-2l}}>=\int\limits_{-\infty}^{+\infty}\frac{dx}
{x^{2n-2k-2l}} =0.
$$
\end{lem}
Here we use our standard rule that the integration contour goes
around zero in the
complex domain, $2k<n$, $2l<n$.

The proof is obvious.

\begin{lem}
\label{lem:l8}
Let us assume, that we have only one singular point  $x_1=0$
of the
order $n$, and $\cN$ is an integer such, that $\cN>n$.

Consider the following collection of functions $\Xi_{l}(x,\varepsilon)$,
$l=0,1,\ldots,\left[\frac{n-1}{2}\right]$ in our space  $\cF_{X}$:
$$
\Xi_{l}(x,\varepsilon)=\frac{1}{\sqrt{\varepsilon}}
\left[\frac{\varepsilon}{x}\right]^{n-2l}  e^{-[x/\varepsilon]^{2\cN}},
$$
The Gram matrix $g_{kl}^{n}$ for this collection of functions
\begin{equation}
\label{eq:l8}
g_{kl}^{n}=<\Xi_{k}(x,\varepsilon),\Xi_{l}(x,\varepsilon)>
\end{equation}
is negative defined and does not depend on $\varepsilon$.
\end{lem}
{\bf Proof.} Consider any linear combination of these functions
$$
f(x)=\sum\limits_{k=0}^{n-1}d_k\Xi_{k}(x,\varepsilon)
$$
We have
$$
<f,f>=\frac{1}{\varepsilon}\int\limits_{-\infty}^{+\infty}
\sum\limits_{k=0}^{n-1}\sum\limits_{l=0}^{n-1} d_k\overline{d_l}
\left[\frac{\varepsilon}{x}\right]^{2n-2k-2l} [e^{-2[x/\varepsilon]^{2\cN}}-1]dx+
$$
$$
+\frac{1}{\varepsilon}\int\limits_{-\infty}^{+\infty}
\sum\limits_{k=0}^{n-1}\sum\limits_{l=0}^{n-1} d_k\overline{d_l}
\left[\frac{\varepsilon}{x}\right]^{2n-2k-2l} dx
$$
The second integral is equal to zero by Lemma~\ref{lem:l7} and the integrand
in the first integral is real, regular and strictly negative, therefore
$$
<f,f>\ <0.
$$
The second statement immediately follows from the scaling properties.

\begin{lem}
\label{lem:l9}
We assume (as in the previous lemma) that we have only one singular point
 at the point $x_1=0$ of the
order $n$, and $\cN$ is an integer such that $\cN>n$. Let us fix an
 interval
$[-L,L]$ where $L$ is either any positive number or $+\infty$.

Consider the following collection of functions $\Xi_{l}(x,\varepsilon)$,
$l=0,1,\ldots,\left[\frac{n-1}{2}\right]$ in our space  $\cF_{X}$:
$$
\Xi_{l}(x,\varepsilon)=\frac{1}{\sqrt{\varepsilon}}
\left[\frac{\varepsilon}{x}\right]^{n-2l}\cdot  e^{-[x/\varepsilon]^{2\cN}}
\cdot \zeta(x),
$$
where
$\zeta(0)\ne0$, $\zeta(x)$ is bounded on the interval $[-L,L]$
and smooth inside it, $\zeta'(0)=\zeta''(0)=\ldots=
\zeta^{(2\cN-1)}(0)=0 $. Define the Gram matrix
$\tilde g_{kl}^{n}(\varepsilon)$ for this collection of functions by
\begin{equation}
\label{eq:l9}
\tilde g_{kl}^{n}(\varepsilon)=\int\limits_{-L}^{+L} \Xi_{k}(x,\varepsilon)
\overline{\Xi_{l}(\bar x,\varepsilon)} dx.
\end{equation}
Then
$$
\tilde g_{kl}^{n}(\varepsilon)\rightarrow g_{kl}^{n}\cdot|\zeta^2(0)| \ \ \mbox{as} \ \
\varepsilon\rightarrow 0,
$$
where $g_{kl}^{n}$ are the scalar products from Lemma~\ref{lem:l8}.
 Here we
use our standard rule, that the integration contour goes around the
singular point $x=0$ in the complex domain.
\end{lem}
{\bf Proof.} Let us make the following substitution: $x=\varepsilon y$.
We have
$$
\tilde g_{kl}^{n}(\varepsilon)=\int\limits_{-L/\varepsilon}^{+L/\varepsilon}
\left[\frac{1}{y}\right]^{2n-2k-2l}\cdot  e^{-2y^{2\cN}}\cdot
\zeta(\varepsilon y) \overline{\zeta(\varepsilon \bar y)} dy =
$$
$$
=\zeta(0) \overline{\zeta(0)}\int\limits_{-L/\varepsilon}^{+L/\varepsilon}
\left[\frac{1}{y}\right]^{2n-2k-2l}\cdot  e^{-2y^{2\cN}}dy+
$$
$$
+\int\limits_{-L/\varepsilon}^{+L/\varepsilon}
\left[\frac{1}{y}\right]^{2n-2k-2l} \cdot
\left[\zeta(\varepsilon y)
\overline{\zeta(\varepsilon \bar y)}-  \zeta(0) \overline{\zeta(0)}
\right] \cdot  e^{-2y^{2\cN}}  dy
$$
The first integral converges to $g_{kl}^{n}|\zeta^2(0)|$ as
$\varepsilon\rightarrow 0$. The pre-exponent in the second integral
is bounded. It
 uniformly converges to 0 at any compact interval. Therefore this integral
converges to 0. The proof is finished.

Consider now the generic ''decaying at infinity'' case. We assume that our
potential has $N$ singular points $x_1$,\ldots,$x_N$ with the
multiplicities
$n_1$,\ldots,$n_N$. Let $\cN=\max(n_1,\ldots,n_N)+1$.

Consider the following collection of functions
$$
\Xi_{lj}(x,\varepsilon)=\frac{1}{\sqrt{\varepsilon}}\cdot
\left[\frac{\varepsilon}{x-x_j}\right]^{n_j-2l}
e^{-\left[\frac{x-x_j}{\varepsilon}\right]^{2\cN}}
\cdot \zeta_j(x),
$$
where
$$
\zeta_j(x)= \left[\prod\limits_{{k\ne j}\atop{k=1,\ldots,N}}
\frac{((x-x_j)^{2\cN}-(x_k-x_j)^{2\cN})^2}{((x-x_j)^{2\cN}-(x_k-x_j)^{2\cN})^2+
1}\right]^{\cN}
$$
$l=0,\ldots,\left[\frac{n_j-1}{2}\right]$, $j=1,\ldots,N$.

\begin{lem}
All functions $\Xi_{lj}$ belong to the space $\cF_{X}$.
\end{lem}
{\bf Proof.} The function $\Xi_{lj}(x,\varepsilon)$ are symmetric  in $(x-x_j)$ if
$n_j$ is even or skew-symmetric in $(x-x_j)$ if $n_j$ is odd. At all other
singular points $x_l$ they have zeroes of order $2\cN\ge n_l+1$. At infinity they
decay exponentially, therefore all conditions are fulfilled.

\begin{lem}
\label{lem:l10}
The scalar products of  functions defined above have the following form
\begin{equation}
\label{eq:l10}
<\Xi_{l_1j_1}(x,\varepsilon) \Xi_{l_2j_2}(x,\varepsilon)>=
g_{l_1l_2}^{n_{j_1}}\cdot\zeta_{j_1}^2(0)\cdot   \delta_{j_1 j_2}+O(1) \ \
\mbox{as} \ \ \varepsilon\rightarrow0.
\end{equation}
where $g_{kl}^{n}$ denotes the Gram matrix defined by (\ref{eq:l8}).
\end{lem}
{\bf Proof.}
Let $j_1\ne j_2$. Assume that $x_{j_1}<x_{j_2}$,  $2L=x_{j_2}-x_{j_1}$.
The product $\Xi_{l_1j_1}(x,\varepsilon) \Xi_{l_2j_2}(x,\varepsilon)$
is regular
in the whole $x$-line.
Consider the following system of intervals in the $x$-line.

$$
\cI_1=]-\infty,x_{j_1}-L], \ \ \cI_2=[x_{j_1}-L,x_{j_1}+L ], \ \
\cI_3=[x_{j_1}+L,x_{j_2}+L ], \ \ \cI_4=[x_{j_2}+L,+\infty[.
$$

Then we have the following estimates:
$$
|\Xi_{l_1j_1}(x,\varepsilon) \Xi_{l_2j_2}(x,\varepsilon)|\le
\frac{\varepsilon^{n_{j_1}+ n_{j_2} -2l_1 -2l_2-1}}{L^{n_{j_1}+ n_{j_2} -2l_1 -2l_2}}
e^{-2[L/\varepsilon]^{2\cN}} \ \ \mbox{for} \ \ x\in \cI_1\cup\cI_4.
$$
$$
|\Xi_{l_1j_1}(x,\varepsilon) \Xi_{l_2j_2}(x,\varepsilon)|\le
\frac{\varepsilon^{n_{j_1}+ n_{j_2} -2l_1 -2l_2-1}[2\cN]^{2\cN}3^{4\cN^2-2\cN}  }{L^{n_{j_1}-2l_1 -4\cN^2}}
e^{-[L/\varepsilon]^{2\cN}} \ \ \mbox{for} \ \ x\in \cI_2
$$
$$
|\Xi_{l_1j_1}(x,\varepsilon) \Xi_{l_2j_2}(x,\varepsilon)|\le
\frac{\varepsilon^{n_{j_1}+ n_{j_2} -2l_1 -2l_2-1}[2\cN]^{2\cN}3^{4\cN^2-2\cN}  }{L^{n_{j_2}-2l_2 -4\cN^2}}
e^{-[L/\varepsilon]^{2\cN}} \ \ \mbox{for} \ \ x\in \cI_3
$$
From these estimates it follows that all integrals exponentially decays as
$\varepsilon\rightarrow0$.

For $j_1=j_2$ the asymptotic formula for the scalar products immediately
follows from Lemma~\ref{lem:l9}.

As a corollary of Lemma~\ref{lem:l10} we immediately obtain, that for sufficiently small
$\varepsilon>0$ the scalar product on our system of functions is
negative defined. It completes the proof for fast decaying case.

\subsection{Periodic case.}

To simplify the formulas below we  assume  that the
 period of our
potential is equal to $\pi$ in this section.

Consider the following collection of functions
$$
\Xi_{lj}(x,\varepsilon)=\frac{1}{\sqrt{\varepsilon}}\cdot
\left[\frac{\varepsilon}{\sin(x-x_j)}\right]^{n_j-2l}
e^{-\left[\frac{\sin(x-x_j)}{\varepsilon}\right]^{2\cN}}
\cdot \zeta_j(x,\varepsilon) \cdot e^{ic_j\alpha(x)}  ,
$$
where
$$
\zeta_j(x,\varepsilon)= \left[\prod\limits_{{k\ne j}\atop{k=1,\ldots,N}}
\frac{([\sin(x-x_j)]^{2\cN}-[\sin(x_k-x_j)]^{2\cN})^2}
{([\sin(x-x_j)]^{2\cN}-[\sin(x_k-x_j)]^{2\cN})^2+
1} \right]^{\cN},
$$
$$
\alpha(x)=\frac{\int\limits_0^{x} \prod\limits_{k=1}^N [\sin(y-x_k)]^{2\cN} dy }{\int\limits_0^{\pi}
\prod\limits_{k=1}^N [\sin(y-x_k)]^{2\cN} dy }
$$
$l=0,\ldots,\left[\frac{n_j-1}{2}\right]$, $j=1,\ldots,N$, the constants $c_j$
are chosen to provide the proper periodicity:
$$
e^{ic_j} = (-1)^{n_j} \varkappa_0.
$$

It is easy to check, that all functions $\Xi_{lj}$ belong to the space
$\cF_{X}$.

\begin{lem}
\label{lem:l19}
The scalar products of the functions defined above have the following form
\begin{equation}
\label{eq:l19}
<\Xi_{l_1j_1}(x,\varepsilon) \Xi_{l_2j_2}(x,\varepsilon)>=
g_{l_1l_2}^{n_{j_1}}\cdot\zeta_{j_1}^2(0)\cdot   \delta_{j_1 j_2}+O(1) \ \
\mbox{as} \ \ \varepsilon\rightarrow0.
\end{equation}

where $g_{kl}^{n}$ denotes the Gram matrix defined by (\ref{eq:l8}).
\end{lem}
{\bf Proof.}

Let $j_1\ne j_2$. Then all products $\Xi_{l_1j_1}(x,\varepsilon)
\overline{\Xi_{l_2j_2}(\bar x,\varepsilon)}$ are regular functions,
 uniformly
exponentially converging to 0 as $\varepsilon\rightarrow 0$.

Let $j_1=j_2$. It is sufficient to introduce a new variable $y=\sin(x)$.
After that we
apply Lemma~\ref{lem:l9}

This completes the proof of the Statement 3.

\section*{Appendix~3}
\addtocounter{section}{1}
\setcounter{subsection}{0}

Let us consider the following Problem: How many poles the real singular
$T$-periodic  finite-gap
1D Schrodinger Operator might have at the period $[0,T]$?
This question is especially interesting for the case
of R-Fourier Transform. We present here a complete answer to this question
for the Riemann surfaces $\Gamma$ with real branching points
$$w^2=(\lambda-E_0)...(\lambda-E_{2g}), \ E_k\in \RR$$
For the smooth periodic operator with the same spectral curve $\Gamma$
we have  spectral zones $$[E_0,E_1],[E_2,E_3],...,[E_{2g-2},E_{2g-1}],
[E_{2g},\infty]$$
Consider the trace of  monodromy matrix $T(\lambda)$ corresponding
to the period $T$. It has $k_j$ maxima and minima strictly inside of the spectral zone
$$[E_{2j-2},E_{2j-1}],\ j=1,\ldots,g, k_j\geq 0$$

\begin{thm}
Let the potential $u(x)$ correspond to the case of R-Fourier
Transform and Riemann surface $\Gamma$
as above.  The total number of singularities at the period $[x,x+T]$
 is equal
to the number
$$n(u)=(k_{g}+1)+(k_{g-2}+1)+\ldots+(k_{g-2i}+1)+\ldots+(k_2+1),\ g=2s$$
$$n(u)=(k_{g}+1)+(k_{g-2}+1)+\ldots+(k_{g-2i}+1)+\ldots+(k_1+1), \ g=2s+1$$
Each singularity has a local form $\sim n_l(n_l+1)/(x-x_l)^2$.
For $x_l=0$ we have singularity with $n_0=g$. There exists only one singular
 point with $n_l=g$, i.e. $n_l<n_0$ for $l\neq 0$. By definition,
 $n(u)=\sum\limits_l n_l$

\end{thm}

 For the famous Lam\'e potentials $u=g(g+1)\wp(x)$ corresponding
 to the rectangular lattice
 we have all $k_j=0$. Our requirement ($n(u)=g$) implies  that $k_j=0$
  for the even
  values of $j+g$ only. So there are many R-Fourier Transform
  potentials of that type (i.e. with
  one pole at the period where the Hermit type Dirichlet Problem
   makes sense).

    {\bf The proof of this theorem is based on the
    results of this work:
  A comparison of the number of negative squares in the inner product
 with the number of poles of potential, and the description of
 the functional
   space $\cF_{X}$ are necessary for the proof.}

  It is easy to classify all realizable collections $k_j$.
The case of Lam\'e type elliptic potentials corresponding to
the rhombic
lattices, as well as more general cases of Riemann surfaces with complex
branching points will be described in the next work.

\begin{rem}
a)The main goal of our work is to develop the theory of the R-Fourier
Transform
which is the best possible analog of Fourier transform
(with good multiplicative properties) on Riemann surfaces.
It should be applied to
 the spaces of smooth
functions defined
in the special ''canonical''contours on the real Riemann surfaces.
 It involves
 quite
original spectral theory for the singular operators on the whole real
$x$-line.
This spectral theory is based on the special indefinite inner products
in the spaces of functions,
containing  singular functions.
We  realized this program for the real singular finite-gap
(i.e. algebro-geometrical) operators,
but it certainly can be extended to the infinite-gap cases also.
The
corresponding operator $L$ is defined in the same space  $\cF_{X}$
as before.
It is symmetric in the same  indefinite inner product. However we did not
proved yet the completeness of the corresponding basis.
 An obstacle for that can
be found in the works \cite{Sk}, \cite{GT},
  \cite{DjakMit}, \cite{DjakMit2}.
The Darboux--Crum transformations play only technical
role here. We do not consider them as a really necessary
part of our
theory.

b) In our work we invented the following class of real potentials with the
special isolated  singularities: for all values of complex spectral
parameter the solutions $\psi(x)$ should be locally meromorphic in the
infinitesimal neighborhood of the corresponding singular points at the
real $x$-axis. We found no classical or modern works  where this
 property had  been
discussed. The analogous property has been used in the work
\cite{GesztWeik}
for the solutions defined in the whole complex $x$-plane
but for the elliptic potentials only: It was used as an assumption
 implying
that such elliptic potential is
algebro-geometrical -- i.e. ``singular finite gap''. In our opinion,
this is the most essential idea
of the work \cite{GesztWeik} related to our theory.

c) There is a problem concerning a completion of our functional
classes. We proved the decomposition theorems for locally smooth functions
 (outside of singularities),
but Hilbert Spaces do not work here.

\end{rem}

\begin{rem}
a)Deconinck and Segur made the following claim in the section 4.4
of their work
\cite{DecSeg}:
According to the KdV dynamics (with hierarchy) the poles of
finite-gap
elliptic, rational and trigonometric solution
can collapse to singular points  by the triangular groups
 containing
$n(n+1)/2$ items only. It  is true, but they claim also
 that the poles are
 leaving this point in the
complex $x$-plane  along the directions of  vertices of the
 equilateral polygon.  It is wrong
for $n>2$.
In fact, according to our results \cite{GrinNov_UMN_2009},
 \cite{GrinNov_DAN_2011}
this multiple pole splits ``approximately'' ( but not exactly)
as a set of integer points in the equilateral triangle for all
$n=1,2,3,\ldots$. Exactly
$[(n+1)/2]$ poles remain at the real axis which is a diagonal in this
triangle. This number plays a fundamental role in our results
\cite{GrinNov_UMN_2009}, \cite{GrinNov_DAN_2011}.
Each real generic pole contributes to the total number
of negative squares in the indefinite metric for which our
  singular Schr\"odinger operator is symmetric.

b)Until now we  cannot understand the proof of the main statement of
 the work \cite{DecSeg}: is it true
 that their
elliptic deformations of multisoliton potentials  are always
finite-gap?
 We are planning to clarify this question later.
\end{rem}


\begin{thebibliography}{99}

\bibitem{GrinNov_UMN_2009} Grinevich, P.G., Novikov, S.P.,
    Singular finite-gap operators and indefinite metrics,
    \textit{Russian Mathematical Survey}, \textbf{64}:4 (2009), 625--650.

\bibitem{GrinNov_DAN_2011} Grinevich, P.G., Novikov, S.P.,
    Singular Solitons and Indefinite Metrics,
    \textit{Doklady Mathematics}, \textbf{83}:3 (2011), 56--58.

\bibitem{DMN76}Dubrovin,B.A., Matveev, V.B., Novikov, S.P.,
    Nonlinear equations of the Korteweg-de-Vries type, finite-zone
    linear operators and  Abelian Varieties,
    \textit{Russiam Mathematical Surveys}, \textbf{31}:1 (1976), 55-136.

\bibitem{Krich_FA_1986} Krichever, I.M.,
    Spectral theory of finite-zone nonstationary Schr\"odinger operators.
    A nonstationary Peierls model. \textit{Functional Analysis and Its Applications},
    \textbf{20}, No. 3,(1986), 203--214.

\bibitem{Krich_UMN_1989} Krichever, I.M.,
    Spectral theory of two-dimensional periodic operators and its applications,
    \textit{Russian Mathematical Surveys}, \textbf{44}:2(266) (1989), 121--184.

\bibitem{String} Krichever, I.M., Novikov, S.P.,
    Riemann Surfaces, Operator Fields, Strings. Analogues of the Laurent-Fourier Bases,
    \textit{Memorial Volume for Vadim Kniznik, ''Physics and Mathematics of Strings''},
    eds L.Brink, E.Friedan, A.M.Polyakov, World Scientific Singapore, (1990), 356-388.

\bibitem{ArkPorgPol} Arkad'ev, V.A., Pogrebkov, A.K.,  Polivanov, M.K.,
    Singular solutions of the KdV equation and the inverse scattering method,
    \textit{Journal of Soviet Mathematics}, \textbf{31}:6 (1985), 3264--3279.

\bibitem{ClMan}  Clarkson, P.A., Mansfield, E.L., The second Painlev\'e equation,
     its hierarchy and associated special polynomials,
     \textit{Nonlinearity} {\bf 16} (2003) R1-R26.

\bibitem{Sk}  Shkalikov, A.A., Veliev, O.A.,  On the Riesz basis property of
the eigen- and associated functions of periodic and antiperiodic
Sturm-Liouville problems. \textit{Mathematical Notes}, {\bf 85},
(2009), Issue 5-6, 647--660.

\bibitem{GT} Gesztesy, F., Tkachenko, V., A criterion for Hill operators to be spectral operators of scalar type. \textit{J. d'Analyse Math.}, {\bf 107},
(2009), 287--353.

\bibitem{DjakMit} Djakov, P., Mityagin, B., Convergence of spectral decompositions
of Hill operators with trigonometric polynomial potentials. \textit{Math. Ann.} {\bf 351}
(2011), no. 3, 509--540.

\bibitem{DjakMit2} Djakov, P., Mityagin, D., Criteria for existence of Riesz bases
consisting of root functions of Hill and 1D Dirac operators.
\textit{J. Funct. Anal.} {\bf 263} (2012), no. 8, 2300--2332.

\bibitem{GesztWeik} Gesztezy, F., Weikard, R., Picard potentials and Hill's equation on a torus.
    \textit{Acta Math.}, \textbf{176} (1996), 73--107.

\bibitem{DecSeg} Deconinck, B., Segur H., Pole Dynamics for Elliptic Solutions of
    the Korteweg-de Vries Equation. \textit{Mathematical Physics, Analysis and Geometry},
    \textbf{3}:1 (2000), 49--74.

\bibitem{DuGr} Duistermaat, J. J., Gr\"unbaum, F. A. Differential equations in the spectral parameter. 
\textit{Communications in Mathematical Physics}, \textbf{103} (1986), no. 2, 177--240.

\bibitem{GrinNov_UMN_2013} Grinevich, P.G., Novikov, S.P.,
Spectral Meromorphic Operators and Nonlinear Systems, 
\textit{Uspekhi Mathematicheskich Nauk}, \textbf{69}:5(419 (2014), 163-164 
(In Russian); doi:10.4213/rm9621;

\bibitem{GrinNov_arxiv_2013} Grinevich, P.G., Novikov, S.P.,
Spectral Meromorphic Operators and Nonlinear Systems,  arXiv:1409.6349.
 

\end{thebibliography}
\end{document}